# Topological Protection in a Landau Flat Band at $v = 7/11$, a Candidate Filling Factor for Unconventional Correlations

Waseem Hussain[1], Haoyun Huang[1], Loren N. Pfeiffer[2], Kenneth W. West[2], Kirk. W. Baldwin[2], and Gábor A. Csáthy[1] *

[1]Department of Physics and Astronomy, Purdue University; West Lafayette, Indiana 47907, USA

[2]Department of Electrical Engineering, Princeton University; Princeton, New Jersey 08544, USA

Corresponding Author:

Gábor A. Csáthy

gcsathy@purdue.edu

Department of Physics and Astronomy, Purdue University; West Lafayette, Indiana 47907, USA.

765-494-3045

## Abstract:

Strong interactions in Landau flat bands are known to stabilize correlated states that do not form in other types of flat bands. We report hallmarks of topological protection at the Landau level filling factor $v = 7/11$ in a two-dimensional electron system. The $v = 7/11$ filling factor is the particle-hole conjugate of $v = 4/11$, a filling factor intensely studied for the possibility of realizing unconventional electronic correlations. Our data establishes a new instance for an unusual fractional quantum Hall state and opens up possibilities for the study of unconventional correlations in an enlarged parameter space. We report and discuss transport signatures developing at other filling factors of interest $v = 11/17, 5/8$ and $8/13$, which however in our sample do not exhibit topological protection.

## Significance:

Flat band systems are currently under intense scrutiny because they support various strongly correlated ground states. Examples are fractional Chern insulators in magic-angle graphene and common fractional quantum Hall states in the 2D electron gas. These two groups of states belong to the same universality class. However, the 2D electron gas is predicted to host more exotic states that are not present in magic-angle graphene. We report a candidate for such an exotic unconventional state at the filling factor ν = 7/11. Our result establishes a novel state with unusual electronic correlations and opens up possibilities for the study of unconventional correlations in an enlarged parameter space.

## Introduction:

Strongly correlated electronic systems continue to play a central role in contemporary condensed matter physics partly because they hold the promise of exotic emergent behavior. Flatband systems are known to bolster electronic interactions and thus they promote strong correlations. Some of the most widely known strongly correlated states are the $v = 1/3$ and $v = 1/5$ fractional quantum Hall states (FQHSs)[1] i.e. the Laughlin states[2]. In addition, there are numerous other FQHSs that develop at sequences of Landau level filling factors of the form $v = p/(2p \pm 1)$ and $v = p/(4p \pm 1)$, where $p$ is an integer. These subsets of FQHSs are collectively referred to as Jain states or conventional FQHSs. Jain sequences and their properties can naturally be explained by the population of a set of discrete and degenerate Λ-levels by composite fermions (CFs), emergent particles formed by binding of electrons to an even number of quantized vortices[3].

One of the most exciting and challenging endeavors in the study of correlated matter is the discovery of states with fundamentally new types of correlations. In the fractional quantum Hall regime such novel states are referred to as unconventional FQHSs. Unconventional FQHSs thus have more intricate electronic correlations than those of the Jain states. Because of highly non-trivial properties and potential uses in topological quantum computation[4,5] among the unconventional FQHSs the non-Abelian candidate states have generated the most interest[6]. There are only a handful of unconventional FQHSs; examples are the Moore-Read Pfaffian and the Read-Rezayi parafermionic candidate states[7,8] and there is experimental evidence for unconventional behavior in Refs.[9-12].

Since the $v = 4/11$ FQHS is not part of the $v = p/(2p \pm 1)$ or the $v = p/(4p \pm 1)$ Jain sequences[13-15], it motivated numerous theoretical studies[16-33]. Within the framework of the composite fermion theory[3], this FQHS can be described by mixed-flavor CFs[13,16]. Recent theoretical work suggests that the nature of the $v = 4/11$ FQHS depends on its spin polarization. For partial polarization, correlations are of the conventional Jain type[16,18,27]. However, a fully spin polarized $v = 4/11$ FQHS has unconventional correlations[22,25,27] described by the Wójs, Yi, and Quinn (WYQ) model[22]. The unconventional topological order of the WYQ state stems from the unusual form of the inter-CF interactions. We are unaware of an analytical construction for the WYQ wavefunction, hence attempts to schematically present the distinction between mixed flavor CF Jain state and WYQ state at $v = 4/11$ are challenging. Instead, the WYQ wavefunction can be calculated using numerical methods[22,25]. The hallmark property of the WYQ wavefunction is the clustering of composite fermions that suppresses the number of CF pairs with a relative angular momentum of three quantum units[22,25]. The onset of this unusual topological order is quantified by a shift invariant distinct from that of a conventional Jain state. The WYQ state may support Abelian[25,29] or non-Abelian[28] quasiparticle excitations. For now, spin polarization and the nature of the topological order of the experimentally realized $v = 4/11$ FQHSs remain unknown and can only be determined by further investigations. However, this FQHS remains a candidate for unconventional topological order and for hosting non-Abelian anyons.

In this work we examine the region of filling factors $3/5 < v < 2/3$ in a modern sample of an exquisite quality by employing ultra-low temperature techniques to cool our sample to a temperature inaccessible in a typical dilution refrigerator. When cooled to 7.6 mK, the weak transport signature at $v = 7/11$ observed at tens of millikelvin develops a full-fledged topological protection. The measured Hall quantization, the nearly vanishing longitudinal resistance, and the opening of an energy gap signal a genuine fractional quantum Hall ground state at $v = 7/11$. This filling factor is the particle-hole conjugate of $v = 4/11$, therefore the FQHS at $v = 7/11$ is a new realization of a highly unusual FQHS. Our work therefore opens up the possibility to study unconventional topological order in an enlarged parameter space. We also observe transport signatures at other nearby filling factors $v = 8/13, 11/17$, and $5/8$ at which, however, in our sample we do not detect topological protection. We discuss these observations from the perspective of particle-hole symmetry.

**Results and Discussion:**

Figure 1. displays the longitudinal magnetoresistance $R_{xx}$ and the Hall resistance $R_{xy}$ over a wide range of magnetic fields as collected at the temperature T=7.6 mK. The thermalization of the sample is achieved through the use of a He$^3$ immersion cell[34] and the mixing chamber temperature is monitored by a carbon thermometer[35] calibrated against a quartz tuning fork He$^3$ viscometer[34]. Data in Fig. 1 are identical to those shown in Fig.1 of Ref.[36]. The quality of the sample is highlighted by the observation of the recently reported reentrant integer quantum Hall state associated with the Wigner solid near $\nu = 1.8$[37] and the fractional quantum Hall state at $\nu = 9/11$ associated with six-flux composite fermions[36].

For the course of this work, we are interested in the magnetic field range between 6.20 T and 6.95 T that falls between the filling factors $\nu = 2/3$ and $3/5$. In this range of magnetic fields Fig.1 exhibits several features. Fig. 2 displays the temperature-dependence of these features over a slightly wider range of magnetic fields. The three top panels Fig. 2a, Fig. 2b, and Fig. 2c show data collected from a cooldown in which the sample was thermalized via the measurement wires. The lowest panel, exhibiting data at 7.6 mK, was generated from data collected on the same sample mounted in a He$^3$ immersion cell. The sample density is different in the two cool-downs by 0.5%. Data in the lowest panel of Fig. 2d is rescaled by this amount in order for the filling factors from different cool-downs to coincide. The successively lower panels of Fig. 2 show data with a decreasing temperature. One effect of the decreasing temperature is the widening of the $\nu = 2/3$ FQHS plateau in these data.

The magnetoresistance $R_{xx}$ shown in Fig. 2a, as measured at T=57 mK, exhibits a rich structure between $\nu = 2/3$ and $3/5$, with several local minima. This contrasts with the earlier observation which exhibited only a single local minimum at $\nu = 7/11$[13]. Fig. 2a also exhibits such a local minimum at $\nu = 7/11$. However, in addition there are three other transport signatures at $\nu = 11/17, 8/13$, and $5/8$. In all, our data at 57 mK exhibit a total of four local minima in $R_{xx}$.

At the lowest temperature of T=7.6 mK transport still exhibits a rich structure. The most salient feature of data at T=7.6 mK is the very deep minimum in $R_{xx}$ we observe at $\nu = 7/11$. At this filling factor, the Hall resistance $R_{xy}$ exhibits a clear quantization to the value of $11h/7e^2$. Furthermore, the temperature dependence of $R_{xx}$ is activated. We indeed find that below the temperature of 77 mK, the magnetoresistance has the $R_{xx} \propto \exp[-\Delta_{7/11}/2k_B T]$ temperature dependence. Such an activated magnetoresistance signals the opening of an energy gap in the spectrum of the magnitude of $\Delta_{7/11}$=21 mK. The value of the energy gap $\Delta_{7/11}$ was extracted by performing a linear fit to the low temperature region of Arrhenius plot shown in Fig. 3. The vanishingly small $R_{xx}$, the quantized $R_{xy}$, and the opening of an energy gap signaled by an activated transport contrast earlier results[13] and are key signatures establishing topological protection and a fractional quantum Hall ground state at $\nu = 7/11$. This state in our experiment is thus incompressible. We note that, as seen in Fig.2, the magnetic field of the local $R_{xx}$ minimum near $\nu = 7/11$ exhibits a minute variation with the temperature. Such an effect may be due to a temperature-dependent resistive background, proximity to a spin transition, or it may be the harbinger of a phase transition of a different nature. However, the value of the quantized Hall resistance allows us to identify the ground state with a fractional quantum Hall state with confidence.

Here we highlight the importance of generating ultra-low temperatures. Data from a sample of similar parameters and similar quality but collected at 30 mK[38], i.e. at a temperature about a factor of 4 higher than the lowest value we achieved, exhibit magnetoresistance features in the same filling factor range, i.e. for $3/5 < \nu < 2/3$. However, these data did not yield new insight.

For further understanding we suggest that the $\nu \leftrightarrow 1 - \nu$ particle-hole symmetry is applicable[39,40]. The $\nu = 7/11$ FQHS is then the particle-hole conjugate of the $\nu = 4/11$ FQHS, or $7/11 \leftrightarrow 4/11$. The $\nu = 7/11$ FQHS therefore may be thought of as a new realization of the $\nu = 4/11$ FQHS. By inheriting properties of the $\nu = 4/11$ FQHS, the $\nu = 7/11$ FQHS is also a candidate for unconventional correlations.

In order to understand the $\nu = 7/11$ FQHS, in the following we review theoretical efforts focused on the $\nu = 4/11$ FQHS, the particle-hole conjugate of the $\nu = 7/11$ FQHS. Jain sequences at $\nu = p/(2p \pm 1)$ and $\nu = p/(4p \pm 1)$ can be accounted by CFs of different flavor, i.e. by two-flux CFs ($^2$CFs) and four-flux CFs ($^4$CFs), respectively[3]. Since 4/11 is located between 1/3 and 2/5, two consecutive members of the $\nu = p/(2p \pm 1)$ Jain sequence, a Jain state is not expected at $\nu = 4/11$. However, a FQHS at $\nu = 4/11$ is allowed by the hierarchy scheme, if quasiparticles of the parent state $\nu = 1/3$ FQHS condense[31-33]. The CF theory provided an elegant rationale for the FQHS at $\nu = 4/11$ when CFs of mixed flavor, i.e. when both two-flux and four-flux CFs, are considered[13,16]. In the first step of this process $^2$CFs form. At $\nu = 4/11$, the lowest $\Lambda$-level of $^2$CFs is completely filled, whereas the second $\Lambda$-level is only filled to one third of its capacity. If the $^2$CFs do not self-interact, a FQHS will not develop at $\nu = 4/11$. However, interactions between the $^2$CFs in the upper $\Lambda$-level may cause their condensation into a FQHS. This process of forming the $\nu = 4/11$ FQHS was dubbed the fractional quantum Hall effect of composite fermions[13]. The condensation process involves the capture of two additional vortices by the $^2$CFs of the upper $\Lambda$-level. The $^4$CFs formed this way generate their own $\Lambda$-levels and, at $\nu = 4/11$, exactly one such $\Lambda$-level associated with $^4$CFs will be filled. As a result of this construction, the mixed-flavor CF state at $\nu = 4/11$ has $^4$CFs in the highest occupied energy level and, in the limit of vanishing residual interaction between the $^4$CFs, it inherits conventional Jain correlations. A significant amount of early theoretical and numerical work supported the idea that the FQHS at $\nu = 4/11$ is a mixed-flavor CF state with conventional Jain correlations[18-21]. A two-component spin singlet state is thought to admit such a description[27]. Furthermore, there is another two-component partially polarized candidate for this FQHS[27]. However, a fully spin polarized $\nu = 4/11$ FQHS has a different origin[25,27].

As an alternative to conventional Jain states, the idea was advanced according to which certain types of inter-CF interactions may stabilize unconventional topological order. For the FQHS at $\nu = 4/11$ it was pointed out by Wójs, Yi, and Quinn that the inter-CF interaction has a highly unusual property of suppressing the number of pairs of CFs with a relative angular momentum of three quantum units[22]. They showed that such interactions confer a novel topological order to the FQHS at $v = 4/11$ that is distinct from that of a mixed-flavor Jain state[22-29]. This difference manifests in a different Wen-Zee topological invariant[22,25] and in an unusual magneto-roton minimum of the WYQ state[26]. The high overlap of the ground state wavefunction with the composite-fermionized WYQ wavefunctions provided evidence that the fully spin polarized $v = 4/11$ FQHS originates from the unconventional WYQ mechanism[22,25,27]. A careful examination of the role of the spin degree of freedom in the formation of the $v = 4/11$ FQHS yielded a detailed spin phase diagram[27]. Following a different approach, Balram proposed an unconventional wavefunction at $v = 4/11$ based on a partonic description[30]. The $4\bar{2}1^3$ partonic state is a viable candidate for the $v = 4/11$ FQHS as it has a pair-correlation function with a good overlap with that of the exact Coulomb ground state and it has a ground state energy that agrees well with the energy of the Coulomb ground state. The Wen-Zee shift of the $4\bar{2}1^3$ partonic state is different from that of the conventional mixed-flavor Jain state but it is equal to that of the WYQ state[30]. To conclude, the most recent theoretical efforts have shown that both the $v = 4/11$ and the $v = 7/11$ FQHSs remain candidates for unconventional correlations.

The $v \leftrightarrow 1 - v$ particle-hole symmetry applies only to single-component FQHSs, such as fully spin polarized FQHSs. Since the nearby $v = 2/3$ FQHS is known to undergo a spin transition[41], particle-hole symmetry applies to this FQHS only when it is fully spin polarized. In the following we use data available from the literature to argue that the $v = 2/3$ FQHS realized in our sample is such a fully spin polarized state. The spin polarization of a FQHS is known to be density dependent, with the fully spin polarized state developing only past a critical sample density. Furthermore, the critical density for a full spin polarization is a function of the width of the wavefunction in the direction perpendicular of the plane of the electron gas[42]. For quantum well samples, the width of the wavefunction scales with the width of the confining quantum well. The critical density at $v = 2/3$ for a 30nm quantum well was reported $1.1 \times 10^{11}/cm^2$ [43] and that for a 65nm quantum well $0.3 \times 10^{11}/cm^2$ [44]. A linear interpolation to 49nm, the width of the quantum well in our sample, gives us an estimate of $0.7 \times 10^{11}/cm^2$ for the critical density. The density of our sample exceeds this estimate, we therefore infer that the $v = 2/3$ FQHS in our sample is fully spin polarized. The presence of a spin polarized $v = 2/3$ FQHS in our sample underscores the $2/3 \leftrightarrow 1/3$ particle-hole symmetry and suggests the relevance of this symmetry in a larger neighboring region of filling factors that includes $v = 7/11$.

Besides the transport signatures at $v = 7/11$, our data collected at T=57 mK, shown in Fig. 2a, exhibits additional features at $v = 8/13, 5/8$, and $11/17$. We also notice that the local minimum at $v = 8/13$ is present in $R_{xx}$ in Fig. 2a, Fig. 2b, and Fig. 2c, but not in Fig. 2d. At the lowest temperature of T=7.6 mK, at $v = 8/13$ we observed a weak signature in $R_{xx}$ that may be described as an inflexion point and that can no longer be described as a local minimum. Thus, our data shown in Fig. 2 suggests that the temperature dependence of $R_{xx}$ at $v = 11/17, 5/8$ and $8/13$ is very different from that at $v = 7/11$. The contrasting temperature dependence of

$R_{xx}$ at three of these filling factors is shown in Fig. 4 at the lowest temperatures $R_{xx}$ monotonically decreases as function of T at $v = 7/11$ and it is monotonically increasing with T at $v = 8/13, 5/8,$ and $11/17$. We find that the temperature dependence of $R_{xx}$ at $v = 8/13, 5/8,$ and $11/17$ is inconsistent with an activated behavior and, therefore, at these filling factors we found no evidence for the presence of an energy gap. Furthermore, the Hall resistance $R_{xy}$ at these filling factors does not exhibit quantization. Therefore, in contrast to the ground state at $v = 7/11$, those at $v = 8/13, 5/8,$ and $11/17$ remain compressible in our experiment and thus topological protection at these filling factors is absent.

Even though topological protection is not present at $v = 8/13, 5/8,$ and $11/17$, we cannot rule out the development of topological protection in future experiments on samples of either improved quality or samples with different growth parameters. Within the framework of the composite fermion theory, transport features at $v = 7/11, 8/13,$ and $5/8$ can be associated with mixed flavor CFs[13], with four-flux CFs in the valence energy band; at $v = 5/8$ there is a pairing of these CFs. Similarly, a FQHS at $v = 11/17$ is mixed-flavor CFs, with six-flux CFs in the valence energy band. At $v = 5/8$ there is a pairing of the two-flux CFs. Recent theoretical efforts generated exotic candidate states at the particle-hole conjugates of these filling factors[27, 45-48].

The $v \leftrightarrow 1 - v$ symmetry links not only the $7/11 \leftrightarrow 4/11$, but all other filling factors of relevance in the $3/5 < v < 2/3$ and $1/3 < v < 2/5$ range. Indeed, $8/13 \leftrightarrow 5/13, 11/17 \leftrightarrow 6/17,$ and $5/8 \leftrightarrow 3/8$. Furthermore, the symmetry related pairs exhibit similar compressibility. For example, ground states of the symmetry-related pairs $2/3 \leftrightarrow 1/3, 3/5 \leftrightarrow 2/5,$ and $7/11 \leftrightarrow 4/11$ are incompressible and they exhibit topological protection. In contrast, ground states of the symmetry-related pairs $8/13 \leftrightarrow 5/13, 11/17 \leftrightarrow 6/17,$ and $5/8 \leftrightarrow 3/8$ are compressible and lack topological protection in existing data. The only exception is the $8/13 \leftrightarrow 5/13$ pair: in our sample the ground state at $v = 8/13$ is compressible, whereas the ground state at $v = 5/13$ in Ref.[15] is incompressible, albeit with a fragile gap.

We now consider a second emergent symmetry of the fractional quantum Hall regime, the so called particle-hole symmetry of CFs[49]. If $v_{CF}$ is the filling factor of the topmost Λ-level, particle-hole symmetry of CFs relates CF filling factors $v_{CF} \leftrightarrow 1 - v_{CF}$ that belong to a given Λ-level. In the second Λ-level, this symmetry translates to relating electronic filling factors of the form $(1 + v_{CF})/(2(1 + v_{CF}) + 1) \leftrightarrow (2 - v_{CF})/(2(2 - v_{CF}) + 1)$. For example, for $v_{CF} = 1/3$ one obtains $4/11 \leftrightarrow 5/13$. In the same way, $7/11 \leftrightarrow 8/13$. It is interesting to note that no magnetotrasport signature was observed at $v = 9/23$, which is the conjugate filling factor of $v = 6/17$ (here $v_{CF} = 1/5$). Similarly, in our experiment we do not measure any signatures at $v = 14/23$, the CF conjugate filling factor of $v = 11/17$. We note, that $v = 9/23$ and $v = 14/23$, at which no transport signature was seen, are related by the *electronic* particle-hole conjugation $9/23 \leftrightarrow 14/23$, rather than CF particle-hole conjugation. This finding further highlights the $v \leftrightarrow 1 - v$ symmetry as a fundamental property that links the $3/5 < v < 2/3$ and $1/3 < v < 2/5$ ranges of filling factors.

As mentioned earlier, the local minimum in $R_{xx}$ at $v = 8/13$ present at higher temperatures disappears by morphing into an inflection point at T=7.6 mK. This observation could be explained by a widening of the plateau of the $v = 3/5$ FQHS due to the localization of CF quasiparticles. It has been shown that at large CF quasiparticle densities, interactions between CF quasiparticles can stabilize exotic CF solids, such as the Wigner solid of CFs[50-59] or the bubble phase of CFs[21,60,61]. We think that the disappearance of the local minimum in $R_{xx}$ at $v = 8/13$ at the lowest temperatures occurs when such an exotic CF solid energetically competes with a FQHS. Much remains unknown about these exotic correlated solids.

The spin of the $v = 7/11$ and $4/11$ FQHSs remains unknown. Our experiment does not probe the spin at $v = 7/11$. Numerical predictions estimate the spin transition at $v = 4/11$ under highly idealized conditions, such as in the limit of zero layer thickness[25,27] and no Landau level mixing[24,25,27]. Data available in the literature on the $v = 4/11$ FQHS is not conclusive on its spin[13] and the observation of a spin transition in this state remains elusive. The $v = 4/11$ FQHS is more robust than the $v = 7/11$ FQHS, therefore investigating the spin nature of the latter is expected to be even more difficult.

For the sake of completeness, we note that there is another symmetry in the fractional quantum hall regime, the $v \leftrightarrow 2 - v$. One filling factor of interest related to $v = 7/11$ via this symmetry is $2 - 7/11$. In our sample this filling factor can be observed in Fig.1 near $B = 3.1$ T. The magnetoresistance in this region does exhibit a structure, which however cannot be associated with a FQHS. Similarly, the filling factor of interest $2 - 4/11$ in our sample falls near $B = 2.6$ T. Near this magnetic field our sample does not exhibit any special features. It will be interesting to see if FQHS will develop under different conditions or sample parameters at these two symmetry-related filling factors.

In summary, our ultra-low temperature magnetotransport measurements in the region between $v = 2/3$ and $3/5$ reveal an unconventional candidate FQHS at $v = 7/11$. Such a FQHS may be stabilized when the interaction between the composite fermions has a form that suppresses the number of pairs of CFs with a relative angular momentum of three quantum units. We found that transport exhibits signatures at other neighboring filling factors of $v = 8/13, 11/17$, and $5/8$, albeit incompressibility is not present at these filling factors. Several aspects of particle-hole symmetry were analyzed. Because of the pervasive presence of particle-hole symmetry we think that in the region of filling factors we studied, the symmetry-breaking three-body CF interactions are likely to be negligible and thus the inter-CF interactions are dominated by two-body terms.

## Methods:

Magnetotransport measurements were performed on an ultra-high quality GaAs/AlGaAs sample from the latest generation of growth[38]. The sample used in this study is the same as the one in

Ref.[36]. The two-dimensional electron gas is confined to a 49nm wide quantum well and has a density of $1.01 \times 10^{11}$/cm$^2$ and low temperature mobility 35 x $10^6$ cm$^2$/Vs. Low frequency transport measurements were performed using a standard lock-in technique at 13 Hz, with a 3nA excitation current. For the lowest temperature data ultra-low temperature techniques are used: the sample is thermalized in a He$^3$ immersion cell[34] and the mixing chamber temperature is monitored by a carbon thermometer[35] calibrated against a quartz tuning fork He$^3$ viscometer[34].

## Acknowledgements:

Measurements performed at milliKelvin temperatures at Purdue were supported by the US Department of Energy Basic Energy Sciences Program under the award DE-SC0006671. The Princeton University portion of this research is funded in part by the Gordon and Betty Moore Foundation's EPiQS Initiative, Grant GBMF9615.01 to Loren Pfeiffer. We thank Sean Myers for sharing his data collected at 7.6 mK.

## Authors Contribution:

W.H. and H.H. performed low-temperature magneto-transport measurements. L.N.P., K.W.W. and K.W.B. contributed resources by producing molecular beam epitaxy-grown GaAs/AlGaAs samples and characterized them. L.N.P. acquired funding for the growth of the samples. W.H and G.A.C. analyzed the data and wrote the paper. G.A.C. supervised the project and secured funding for the magnetotransport measurements.

## Competing Interests:

## Data availability:

## Figures:

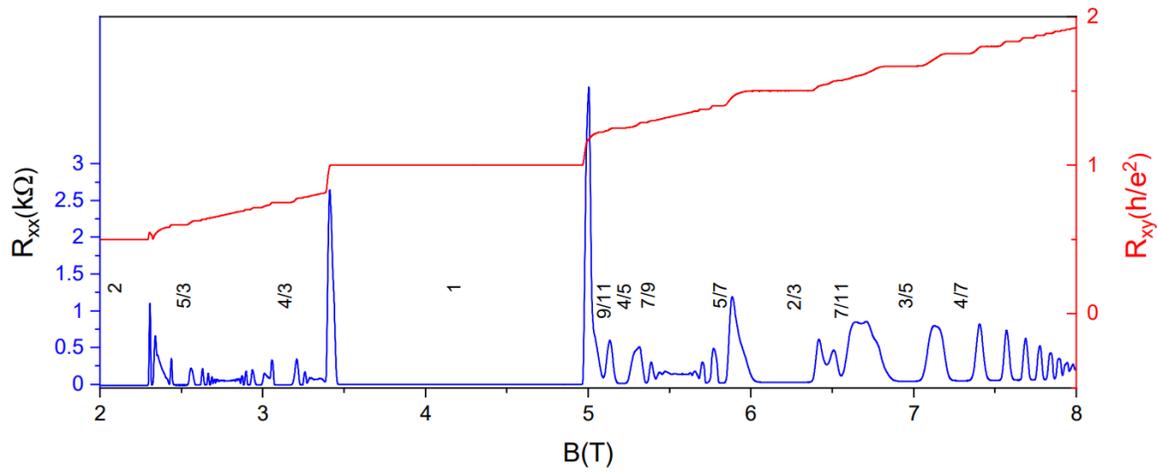

**Figure 1**: Longitudinal magnetoresistance $R_{xx}$ and Hall resistance $R_{xy}$ as plotted against the applied magnetic field B for the region of the lowest Landau level. Data were taken at 7.6 mK. Several Landau level filling factors of interest are marked, including $\nu = 7/11$.

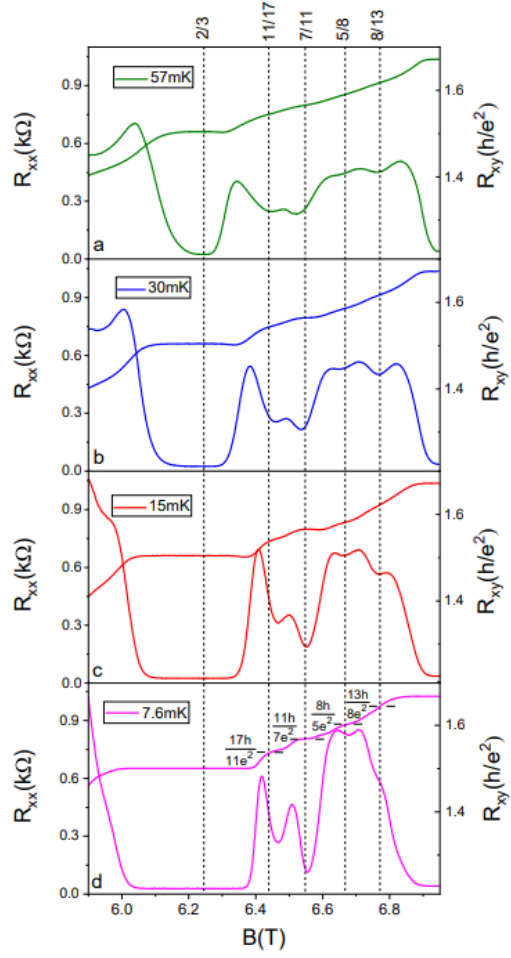

**Figure 2**: The magnetoresistance and the Hall resistance in the magnetic field range between 5.9 T and 6.95 T as measured at four selected values of the temperature. Local minima in $R_{xx}$ are observed at the filling factors $\nu = 7/11, 8/13, 11/17,$ and $5/8$.

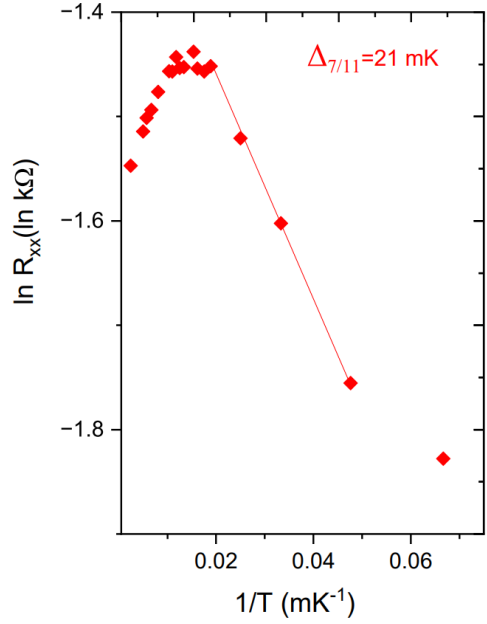

**Figure 3**: Arrhenius plot for the computation of energy gap. The dotted line is a linear fit to the data at the lowest temperatures. The extracted energy gap is $\Delta_{7/11}= 21$ mK.

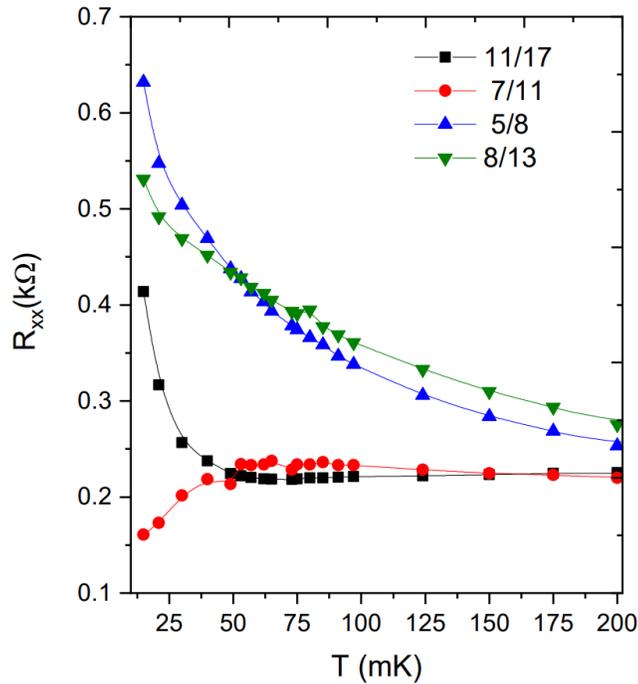

**Figure 4**: The temperature dependence of the magnetoresistance $R_{xx}$ at four selected values of the Landau level filling factor. Lines are guides to the eye. At the lowest temperatures, $R_{xx}$ measured at $\nu = 7/11$ decreases as function of T. In contrast, $R_{xx}$ measured at the other the filling factors increases when the temperature is lowered.